\documentclass[prd,preprint,showpacs]{revtex4}
\usepackage{amssymb,amsmath}
\usepackage{graphicx}% Include figure files

\begin{document}
%\draft

\title{Anomalous post-newtonian terms and the secular increase of the astronomical unit}
\author{L. Acedo}
\thanks{Tel.:+34-963-877007/88285; e-mail: luiacrod@imm.upv.es\,.}
\affiliation{Instituto Universitario de Matem\'atica Multidisciplinar, Building 8G, 2$^\circ$ floor,
Universitat Polit\`ecnica de Val\`encia, Camino de Vera s/n, E-46022, Valencia, Spain}

\date{\today}

\begin{abstract}

In the last decade a major debate has emerged on the astrophysics community concerning the anomalous behaviour of the astronomical unit, the fundamental scale of distances in the Solar system. Several independent studies have combined radar ranging and optical data from the last four decades to come to the conclusion that the astronomical unit is increasing by several meters per century. It is abundantly clear that General Relativity cannot account for this new effect, although an still undefined angular momentum transfer mechanism could provide the simpler and more
conventional explanation. Here we investigate several anomalous post-newtonian terms containing only the product of the mass and angular momentum of the Sun as well as its Schwarzschild radius in order to determine if they could explain the secular increase
of the astronomical unit and the recently reported increase of Lunar's eccentricity. If these anomalies are confirmed, searching for a modification of General Relativity predicting these terms could have far-reaching consequences.

\end{abstract}


\maketitle

\section{Introduction}
\label{sect_1}

\label{intro}

Physics is a science whose final objective is the understanding of the underlying reality to the phenomena. To achieve this objective physical theories are proposed and experiments are developed in order to test them. Accepted theories are those which connect many different phenomena and explain all the experiments and observations in their domain of application. However, it is a well-known fact of the history of science that, from time to time, anomalous observations are gathered or experiments are performed which cannot be explained by any of the theories available at the time. The consequences of these anomalies can be far-reaching: new models could be necessary within the context of existing theories or, eventually, some new physics
must be developed.

In any case, the first step towards the understanding of a new phenomenon is usually descriptive. By means of a pattern unveiled in the data, an empirical law can be formulated. Only later on, this law is explained in the context of a consistent theory.

As any General Relativity student knows, the paradigmatic example of an anomaly in astronomical observations was the excess in the perihelion shift of Mercury that cannot be explained by Newtonian perturbations of the planets.
Conventional explanations were suggested at the end of the nineteenth century and the beginning of the twentieth, usually invoking some extra mass between Mercury and the Sun. Finally, the explanation of this anomaly become after the revolution of the General Theory of Relativity \citep{Roseveare}. The role of the perihelion shift both as a test and a goal for the preliminary versions of the theory cannot
be underestimated \citep{Janssen}. Nevertheless, evidence for anomalies must be very carefully analysed within the context of standard physics before embarking in speculative theories. Precisely because of the crucial role of these anomalies, the search for any possible conventional explanation must be carried out
carefully.

Nowadays, spacecraft missions provide a very stringent test to our understanding of gravitation on the Solar System scale. On the contrary to natural planet and satellites, spacecraft's mass, thermal properties, geometry, etc$\ldots$ are very well-known because of its design. Thanks to the careful monitoring of these missions new effects on spacecraft navigation have been discovered. For example, analysis of the Doppler data for the Pioneer missions to Jupiter and Saturn revealed an 
anomalous acceleration $a­_{P}=(8.74 \pm 1.33) \times 10^{-10}\mbox{ m}/\mbox{s}^2$ directed towards the Sun \citep{Turyshev2010}. 
Many possible conventional and unconventional explanations were proposed \citep{Turyshev2010,Lammerzahl2006}.
However, a very recent study of the whole dataset for the Pioneer 10 and Pioneer 11 orbits strongly suggested that this sunward acceleration is the consequence of anisotropic emission of on-board heat because it decays with time as
expected from a radiative origin \citep{Turyshev2011}. This was discovered after the retrieval of the data for the whole mission and the detailed analysis of finite-element thermal models for the spacecrafts confirmed
this conclusion \citep{Rievers,Turyshev2012}. 

This is not the only unexplained anomaly in Solar System dynamics that have appeared in the last decade: Firstly, as reported by Krasinsky and Brumberg in 2004, the Astronomical Unit, the fundamental distance scale in the Solar System, increases by $15 \pm 4$ meters per century \citep{Krasinsky}. Including more recent measurements by JPL, Standish gave the corrected value of $7 \pm 2$ meters per century \citep{Standish}. 

Since the time of these works, a growing interest on this phenomenon has been building up and many researchers have investigated some theoretical attempts to explain it either conventionally or resorting to new physics. Already in 2005,
Iorio showed that the effect could arise in the Dvali-Gabadadze-Porrati multidimensional braneworld scenario
\citep{Iorio2005}, initially proposed as a way of explaining away accelerated expansion in the Universe. However, DGP cosmology has been criticized
because it does not agree with recent cosmological observations on baryon acoustic oscillations and the cosmic microwave background \citep{Fang2008}. Li and Chang proposed an explanation based upon the kinematics in Finsler geometry \citep{Finsler} while Arakida
has studied and dismissed the effect of cosmological expansion as derived from McVittie spacetime for Schwarzschild geometry embedded in a Friedmann-Lema\^{\i}tre-Robertson-Walker metric \citep{Arakida2011}. The resulting contribution to the increase of the 
semi-major axis of orbits in the Solar system is $9$ orders of magnitude smaller than the observations. 
The most probable conventional explanation to date was discussed by Miura et al. These authors studied the possibility for this expansion of the Solar system to be the consequence of the angular momentum transfer from the Sun to the planets in a 
way similar to the expansion of the Lunar orbit related to tidal friction in the Earth \citep{Miura}. On the other hand, the authors do not describe a concrete mechanism accounting for the angular momentum transfer and they merely calculate the increase
of Solar rotational period in $21$ ms per century.

Another anomaly in the Solar system has been discussed by Anderson and Nieto \citep{Anderson2010,IorioMNRAS2011} on the base of the detailed orbital analysis of William
and Boggs \citep{WilliamsBoggs}. This analysis revealed an anomalous increase in the eccentricity of the Moon's orbit given by $(0.9 \pm 0.3)\times 10^{-11}$ per year. Physical dissipative processes in the Earth
and the Moon are sufficiently well modelled to exclude this rate as an effect of the tidal interaction.

A different approach to these anomalies has been followed by Iorio \citep{IorioAJ2011} who has proposed an empirical model based on the extra radial acceleration:

\begin{equation}
\label{Ioriomodel}
\mathbf{A}_{\mbox{pert}}= k H_0 v_r \mathbf{\hat{r}}\; ,
\end{equation}

where $H_0=7.47 \times 10^{-11} \mbox{yr}^{-1}$ is the Hubble constant at the present epoch, $v_r=d r/d t$ is the radial velocity of the planet or satellite towards the central body and $k$ is a constant of the order of unity. This perturbation predicts the Astronomical unit increase and the Lunar eccentricity anomaly for $2.5 \leq k \leq 5$. The origin of this extra acceleration could be cosmological but it could also arise from a different mechanism with the prefactor
$H_0$ being a numerical coincidence. Anyway, it cannot be presently derived from a theoretical model.

In this paper we consider several anomalous force terms in the context of a perturbation approach to the weak field dynamics of the Solar system. We notice that, in order to contribute to the increase of the semi-major axis of a planetary orbit, a radial force perturbation should be proportional to an odd power of the radial velocity. It could also contain the Newtonian acceleration $\mu / r^2$ as a prefactor, because this is the only acceleration scale in the gravitational problem, or a time scale derived from the rotational angular momentum of the central body. Tangential forces proportional to the tangential velocity are also analysed. Our objective is to
determine what terms could explain both anomalies, parsimoniously, and what we could dismiss. The search for this pattern is not a purely ``ad hoc'' effort because, if one of these anomalous force laws is supported by increasingly accurate observational data, it could prove useful as a methodological guide for a theory.

The paper is organized as follows: In Sec. \ref{sec:2} we evaluate the perturbation
effects on the semi-major axis and the eccentricity of the orbit of a body orbiting
a much larger mass for terms proposed on the basis of dimensional analysis. Three
different kind of perturbations are considered: (i) radial perturbations involving the
mass of the central body (ii) tangential perturbations proportional to the central mass and (iii) radial perturbations proportional to the angular momentum of the central body. We conclude that the last case is the most adequate for a parsimonious 
explanation of the secular increase of the astronomical unit and the Lunar eccentricity. 

\section{Anomalous post-Newtonian terms and Solar system anomalies.}
\label{sec:2}

It is a well-known fact that General Relativity does not predict any significant variation of the semi-major axis in the two-body problem. Only for very massive bodies, orbiting at very close distance, the semi-major axis should decrease as a consequence of the emission of gravitational waves. This prediction has indeed been verified for the Hulse-Taylor system of neutron stars and constitutes the best indirect evidence on the existence of gravitational radiation \citep{HulseTaylor}.

An increase of the magnitude of the astronomical unit disclosed by recent analysis of radar ranging and optical data for the Solar System cannot be accommodated in the context of General Relativity. This does not imply that a conventional explanation
is not possible, but it clearly limits the possible sources for such a behaviour coming from known physics.

In this paper we should assume that a conventional explanation is not possible and that new physics is necessary for a prediction of these phenomena. Our modest objective is to consider the set of perturbation terms that could be added
to the equations of motion of a test particle in a gravitational field in order to explain the anomalous secular increases of the Astronomical Unit and the Moon's orbital eccentricity, without ruining the agreement with classical tests of GR, specially the secular increase of Mercury's perihelion. If these models are logically possible and a numerical agreement with the recently discovered anomalies is achieved, we could think that there is a compelling reason for pursuing extensions
of General Relativity or alternative models.

In proposing these terms we will be guided only by dimensional analysis and the assumption that the origin of these anomalies is not cosmological. This hypothesis restricts the constants to appear in the anomalous force law to: (i) $\mu = G M$, i. e., the product of the gravitational constant and the mass of the central body ,
(ii) Schwarzschild radius of the central body, $r_S= 2 G M/c^2$ (iii) the total angular momentum of the central body corresponding to the rotation around its axis, $J=2 M R^2/5$. The radial orbital velocity, $v_r= \dot{r}$ and the tangential velocity $v_\theta= r \dot{\theta}$ (where the dot denotes derivation with respect to time) should also appear because, as it is clear from the classical theory of perturbations
in celestial mechanics, that only velocity dependent terms could induce secular variations of the semi-major axis of the orbit in the two-body problem.

We will consider two possible perturbation forces: those proportional to $\mu$ and those proportional to the total angular momentum, $J$.  We 
could also have radial or tangential perturbations. The magnitude of the effects is analysed below:

\subsection{Radial perturbations proportional to the Sun's mass}
\label{sec:2:1}

A general term of this kind is given by:

\begin{equation}
\label{radialm}
\delta \mathbf{F}_r(n,m)=\beta_{n,m} \displaystyle
\frac{\mu}{r^2}\left(\displaystyle\frac{r_S}{r}\right)^n \left(\displaystyle\frac{\dot{r}}{c}\right)^m \mathbf{\hat{r}} \; ,
\end{equation}
where $n$, $m$ are positive integers (not simultaneously zero), $\beta_{n,m}$ is a constant
and $\mathbf{\hat{r}}$ is the unit radial vector. The time derivative of the semi-major axis can be obtained from perturbation
theory as follows \citep{Danby,Burns}:
\begin{equation}
\label{dadt}
\displaystyle\frac{d a}{d t}=\displaystyle\frac{2 a^2}{\mu} \dot{E} \; ,
\end{equation}
where $\dot{E}$ is the time derivative of the total energy. If only a radial perturbation force is acting upon the planet
we have $\dot{E}= \dot{r} \delta F_r$ and Eq.\ (\ref{dadt}) can be written as follows:
\begin{equation}
\label{dadt2}
\displaystyle\frac{d a}{d t}=2 \beta_{n,m} a^2 \displaystyle\frac{r_S^n}{r^{n+2} c^m} \dot{r}^{m+1} \; ,
\end{equation}
From the equations of the unperturbed elliptical orbit we have that the orbital radius, $r$, and the radial
velocity, $\dot{r}$, are given in terms of the true anomaly, $\theta$, by:
\begin{eqnarray}
\label{idealorbit}
r &=& \displaystyle\frac{p}{1+\epsilon \cos \theta} \; , \\
\noalign{\smallskip}
\dot{r}&=&\displaystyle\frac{H}{p} \epsilon \sin \theta \; ,
\end{eqnarray}
where $p=a (1-\epsilon^2)$ is the semi-latus rectum of the elliptical orbit, $a$ is the semi-major axis, $\epsilon$ is the
eccentricity and $H=\sqrt{\mu p}$ is the orbital angular momentum per unit mass \citep{Valtonen}. In order to express the derivative with
respect to time in a derivative in terms of the true anomaly we must also use the following relation:
\begin{equation}
\label{dtdth}
d t = \displaystyle\frac{T}{2 \pi} \displaystyle\frac{(1-\epsilon^2)^{3/2}}{(1+\epsilon \cos\theta)^2} d\theta\; ,
\end{equation}
where $T$ is the orbital period. Another useful relation is Kepler's third law given by $\mu= a^3 (2 \pi/T)^2$.
By direct substitution of Eqs.\ (\ref{idealorbit}) and (\ref{dtdth}) into Eq.\ (\ref{dadt}) and making use of the
relations for the ideal elliptical orbit we finally arrive at the expression for the average secular variation
of $a$ in an orbital period, $\langle \dot{a} \rangle = \Delta a/T$, $\Delta a$ being the total variation in the time $T$:
\begin{equation}
\label{aradial}
\langle \dot{a} \rangle_{n,m}= 2 \beta_{n,m} \displaystyle\frac{a}{T} \left( \displaystyle\frac{r_S}{a} \right)^n \left( \displaystyle\frac{2 \pi a}{c T} \right)^m \displaystyle\frac{ \epsilon^{m+1}}{(1-\epsilon^2)^{n+m/2+1}} \mathcal{C}_{n,m} \; ,
\end{equation}
where the coefficients $\mathcal{C}_{n,m}$ are given by the integral:
\begin{equation}
\label{cnm}
\mathcal{C}_{n,m}=\displaystyle\int_0^{2 \pi} \, \sin^{m+1} \theta \left(1+\epsilon \cos \theta \right)^n \, d \theta \; .
\end{equation}
Notice that $\mathcal{C}_{n,m}=0$ for any even integer $m$. A non-zero secular variation of the semi-major axis is only
possible for perturbation forces containing odd powers of the radial velocity. The most simple cases correspond to
$\mathcal{C}_{0,1}=\mathcal{C}_{1,1}= \pi$, $\mathcal{C}_{0,3}=3 \pi/4$. On the other hand, it is expected for Eq.\ (\ref{aradial}) to predict a very small secular increase because it is proportional to powers of the ratio of the Schwarzschild radius of the Sun to the semi-major axis of the Earth orbit, $r_S/a\simeq=1.97 \times 10^{-8}$ as well as to the semi-major axis expressed in light-years, $a/(c T) \simeq 1.58\times 10^{-5}$.

As the increase of the astronomical unit has been deduced from radar ranging data for
all the inner planets it seems adequate to perform an average of Eq.\ (\ref{aradial}) for Mercury, Venus, the Earth and Mars. Semi-major axis, orbital periods and orbital eccentricities are tabulated in standard databases \citep{NASAfactsheet}. For the Sun we also have
$\mu=132,712,440,018$ km$^3/$s$^2$ and the Schwarzschild radius $r_S=2.953$ km.
For the most relevant cases we have:
\begin{eqnarray}
\label{aradresult}
\langle \dot{a} \rangle_{n=0, m=1}&=& \beta_{0,1}\, 6.82 \times 10^8 \;\mbox{cm$/$yr} \; ,\\
\noalign{\smallskip}
\langle \dot{a} \rangle_{n=1, m=1}&=& \beta_{1,1}\, 37.5 \; \mbox{cm$/$yr} \; , \\
\noalign{\smallskip}
\langle \dot{a} \rangle_{n=0, m=3}&=& \beta_{0,3}\, 0.29 \; \mbox{cm$/$yr} \; , 
\end{eqnarray}
For $0.13 \le \beta_{1,1} \le 0.24$ and $ 17.2 \le\beta_{0,3}\le 31.0$ the last two
cases could explain the secular increase of $7\pm2$ cm$/$yr compatible with the last
proposal of Standish \citep{Standish}. But for these values the prediction of the secular increase of 
the Lunar eccentricity is by far too small for our objective.

If we consider only radial perturbations there is a relation among the secular variation of the eccentricity, $\epsilon$, and the secular increase of the semimajor
axis of the orbit as follows:
\begin{equation}
\label{eradial}
\langle \dot{\epsilon} \rangle_{n,m}=\displaystyle\frac{1-\epsilon^2}{2 a \epsilon} \langle \dot{a} \rangle_{n,m} \; .
\end{equation}
We should notice that if the semi-major axis of the elliptical orbit increases then Eq.\ (\ref{eradial}) implies that the orbital eccentricity also increases secularly. This has been also noticed in the model proposed by Iorio \citep{IorioAJ2011}. In the case of the Moon's orbit, the Earth is central body with $\mu=398600.4$ km$^3/$s$^2$. Lunar eccentricity is 
$\epsilon=0.0549$ and the semimajor axis of the orbit $a=384400$ km. From Eqs.\ (\ref{aradial}) and (\ref{eradial}) we obtain:
\begin{eqnarray}
\label{aradresultb}
\langle \dot{\epsilon} \rangle_{n=0, m=1}&=& \beta_{0,1}\, 1.57 \times 10^{-5} \; \mbox{yr$^{-1}$}\; , \\
\noalign{\smallskip}
\langle \dot{\epsilon} \rangle_{n=1, m=1}&=& \beta_{1,1}\, 1.82 \times 10^{-16} \; \mbox{yr$^{-1}$}\; ,\\
\noalign{\smallskip}
\langle \dot{\epsilon} \rangle_{n=0, m=3}&=& \beta_{0,3}\, 1.04 \times 10^{-19} \; \mbox{yr$^{-1}$} \; . 
\end{eqnarray}
Consequently, predictions for the anomalous secular increase of the eccentricity
of the Moon's orbit are either too large or too small. For other values of $n$,$m$ results are even smaller. We conclude that a force term of the form in Eq.\ (\ref{radialm}) could explain the astronomical unit anomaly but not the excess in Lunar's eccentricity increase not explained by known mechanism of angular momentum transfer.

\subsection{Tangential perturbations proportional to the Sun's mass}
\label{sec:2:2}

A similar expression to that given in Eq.\ (\ref{radialm}) can be given for general tangential perturbations:
\begin{equation}
\label{tangentialm}
\delta \mathbf{F}_\theta(n,m)=\beta_{n,m} \displaystyle
\frac{\mu}{r^2}\left(\displaystyle\frac{r_S}{r}\right)^n \left(\displaystyle\frac{r\dot{\theta}}{c}\right)^m \hat{\theta} \; ,
\end{equation}
where $\theta$ is the true anomaly. Energy derivative respect to time is:
\begin{equation}
\label{Edevtime}
\displaystyle\frac{d E}{d t}=r \dot{\theta} \delta \mathbf{F}_\theta(n,m) \; .
\end{equation}
By using Eqs.\ (\ref{idealorbit}) and (\ref{dtdth}) we can calculate the average
increase of energy per revolution as follows:
\begin{equation}
\label{avgE}
\langle \dot{E} \rangle_{n,m}=\beta_{n,m}\displaystyle\frac{\mu}{a T} \left(\displaystyle\frac{r_S}{a}\right)^n \left(\displaystyle\frac{2 \pi a}{c T} \right)^m \displaystyle\frac{1}{(1-\epsilon^2)^{n+m/2+1}} \mathcal{F}_{n+m+1}\; ,
\end{equation}
where
\begin{eqnarray}
\label{Fnm}
\mathcal{F}_N&=&
\displaystyle\int_0^{2 \pi} \, \left(1+\epsilon \cos \theta \right)^N \, d \theta  \\ 
\noalign{\smallskip}
&=&2 \pi \left(1+\pi^{-1/2}\displaystyle\sum_{k=2}^N \, \binom{N}{k} 
\epsilon^{2 k} \displaystyle\frac{\Gamma(k+1/2)}{\Gamma(k+1)}\right) \; .
\end{eqnarray}
For tangential perturbations the orbital angular momentum, $H$, also varies according
to \citep{Burns} :
\begin{equation}
\label{dHdt}
\displaystyle\frac{d H}{d t}=r \delta \mathbf{F}_\theta(n,m) \; .
\end{equation}
The average relative variation of angular momentum is deduces from Eqs. (\ref{tangentialm}), (\ref{idealorbit}) and (\ref{dtdth}):
\begin{equation}
\label{avgH}
\left\langle \displaystyle\frac{\dot{H}}{H} \right\rangle_{n,m}= \beta_{n,m}
\displaystyle\frac{1}{T} \left(\displaystyle\frac{r_S}{a}\right)^n \left(\displaystyle\frac{2 \pi a}{c T} \right)^m \displaystyle\frac{1}{(1-\epsilon^2)^{n+m/2}} \mathcal{F}_{n+m-1}\; .
\end{equation}
From the relation among the time derivative of the semimajor axis and the variation
of total energy in Eqs. (\ref{dadt}) and (\ref{avgE}) we find for this case
a secular increase of $a$ as follows:
\begin{equation}
\label{adevtan}
\langle \dot{a} \rangle_{n,m}=2 \beta_{n,m} \displaystyle\frac{2 a}{T} \left( \displaystyle\frac{r_S}{a} \right)^n \left( \displaystyle\frac{2 \pi a}{c T} \right)^m \displaystyle\frac{1}{(1-\epsilon^2)^{n+m/2+1}} \mathcal{F}_{n+m+1} \; .
\end{equation}
In the tangential perturbation case the variation of eccentricity also depends upon
the secular variation of the orbital angular momentum \citep{Burns}
\begin{equation}
\displaystyle\frac{d \epsilon}{d t}=\displaystyle\frac{\epsilon^2-1}{2 \epsilon}
\left( \displaystyle\frac{\dot{E}}{E}+2 \displaystyle\frac{\dot{H}}{H}\right) \; .
\end{equation}
Therefore, from Eqs. (\ref{avgE}) and (\ref{avgH}) we find:
\begin{equation}
\label{edevtan}
\left\langle \dot{e} \right\rangle_{n,m} = \beta_{n,m} \displaystyle\frac{1}{\epsilon T} \left( \displaystyle\frac{r_S}{a} \right)^n \left( \displaystyle\frac{2 \pi a}{c T} \right)^m \displaystyle\frac{1}{(1-\epsilon^2)^{n+m/2}}\left( \mathcal{F}_{n+m+1}-
(1-\epsilon^2) \mathcal{F}_{n+m-1}\right) \; .
\end{equation}
The most interesting results are obtained for $n=1$, $m=2$ which, from
Eqs. (\ref{adevtan}) and (\ref{edevtan}), are $\langle \dot{a} \rangle=1.3$ cm$/$yr
for Mercury ($\langle \dot{a} \rangle=0.037$ cm$/$yr for the Earth). For $\beta_{1,2}
\approx 10$ this could agree with the astronomical unit anomaly. However, the corresponding increase on Lunar eccentricity is negligible: $\left\langle \dot{\epsilon} \right\rangle = 4.37 \times 10^{-21}$ yr$^{-1}$.

\subsection{Perturbations involving the rotational angular momentum of the central body}
\label{sec:2:3}

A different kind of force term can be constructed from the rotational angular 
momentum per unit mass of the central body, i. e., $J=2/5 \Omega R^2$, where $\Omega$
is the rotation velocity and $R$ is the radius of the Sun or the Earth (depending on
the problem). The most simple small perturbation acceleration involving $J$, the
Schwarzschild radius $r_S$ and the radial velocity of the orbiting body, $\dot{r}$ 
is easily found from dimensional analysis:
\begin{equation}
\label{Jpert}
\delta \mathbf{F}=\beta \Omega \dot{r} \left(\displaystyle\frac{R}{r}\right)^2
\left(\displaystyle\frac{r_S}{r}\right) \mathbf{\hat{r}}\; ,
\end{equation}
with $\beta$ as a constant of the order of unity, as before. Notice that the factor multiplying the radial velocity has the dimensions of an inverse of time:
\begin{equation}
\label{Hr}
H(r)=\Omega \left(\displaystyle\frac{R}{r}\right)^2
\left(\displaystyle\frac{r_S}{r}\right)\; .
\end{equation}
If we now take into account that the rotational period for the Sun is $T=25.05$ days,
its radius, $R=696,342$ km, the Schwarzschild radius $r_S=2.95$ km, and the average
distance from the Earth, $r=149.6*10^5$ km we find that $H(r)=3.92*10^{-11}$ years$^{-1}$ which is very similar to Hubble constant arising for the expansion of the Universe.

Similarly, for the Earth-Moon system we have $T=23.93$ hours for the average sidereal day,
$R=6371$ km for the mean radius, $r_S=8.87$ mm for the Schwarzschild radius and
$r=384,400$ km for the average distance to the Earth. This yields $H(r)=1.45*10^{-11}$
years$^{-1}$, a value which, by chance, is also of the same order of magnitude than 
Hubble constant. This means that the model in Eq. (\ref{Jpert}) is numerically equivalent to Iorio's proposal in Eq. (\ref{Ioriomodel}), a model which, in turn, successfully encompasses the Astronomical Unit and Lunar eccentricity anomalies. However, the interpretation of the model in Eq. (\ref{Jpert}) is very different because it is only related to quantities corresponding to the Solar System.

Following the procedure in Section \ref{sec:2:1} the calculation of the average
increase of the semi-major axis and the eccentricity of the orbit is straightforward
and we finally have:
\begin{eqnarray}
\label{adotJpert}
\left\langle \dot{a} \right\rangle &=& \beta \Omega \left(\displaystyle\frac{R}{r}
\right) \left(\displaystyle\frac{r_S}{a} \right) \, \displaystyle\frac{\epsilon}{(1-\epsilon^2)^{3/2}} a \\
\noalign{\smallskip}
\left\langle \dot{\epsilon} \right\rangle &=& \beta \Omega \left(\displaystyle\frac{R}{r}
\right) \left(\displaystyle\frac{r_S}{a} \right) \, \displaystyle\frac{\epsilon}{2 (1-\epsilon^2)^{3/2}} \; .
\end{eqnarray}
We can check that for the parameters corresponding to the Earth and $\beta=30.5$ we
find that the secular increase of the semi-major axis is, according to Eq.\
(\ref{adotJpert}), $\langle \dot{a} \rangle=5$ cm per year (in the lower bound of
the currently accepted confidence interval). Meanwhile, the Lunar eccentricity increases secularly by $\langle \dot{\epsilon} \rangle=12\times 10^{-12}$ per year.
This value is also within the observed range for the anomaly. It is remarkable that
both anomalies can be deduced from the same perturbation term.

There is still another force term that can be proposed from dimensional analysis. In this case the square of the angular momentum of the central body appears. The most simple term of this form can be written as follows:
\begin{equation}
\label{Jpert2}
\delta \mathbf{F}=\beta \Omega^2 \left(\displaystyle\frac{R}{r}\right)^4
r_S \left(\displaystyle\frac{\dot{r}}{c}\right) \mathbf{\hat{r}}\; ,
\end{equation}
but the prefactor of $\dot{r}$ is $\propto 10^{-18}$ and, consequently, seven orders
of magnitude smaller than the one in Eq.\ (\ref{Jpert}). It should predict an effect too small to account for the anomalies.

\section{Conclusion, remarks and prospects for future work}
\label{sec:3}

Due to the inherent weakness of gravity, General Relativity is not tested with
the same degree of accuracy than the theories of the other forces in nature. After many decades since its proposal, only the two classic tests in the Solar system where known, i.e., deflection of the light by the Sun and the precession of the perihelion
of Mercury \citep{Rindler}. Since the 70's of the past century progress on the confirmation of the
theory has been steadily made. Shapiro effect \citep{Shapiro1964,Shapiro1968}, the Pound-Rebka experiment
\citep{PoundRebka1959,PoundRebka1960}, Hafele-Keating experiment \citep{HafeleKeating}, Gravity probe A for time delay and gravitational redshift \citep{GravityProbeA}, perihelion precession for other planets and the test of the geodetic and Lense-Thirring precession for orbiting gyroscopes performed recently by Gravity Probe B \citep{GravityProbeB}.
Outside the Solar system, the Hulse-Taylor binary pulsar provided an unmistakably clear indirect test on the emission of gravitational radiation in very massive and
close binary stars \citep{HulseTaylor}. However, despite these successes in the experimental confirmation
of the theory, there is still the possibility of smaller effects of deviation from 
the predictions that could only be unveiled with experiments of higher accuracy.
The Astronomical Unit and Lunar eccentricity anomalies could be examples of two effects that would require some new physics beyond General Relativity. 

It is a testament to the status of General Relativity as a provisional, although
very successful theory, the fact that, since its very inception, there have been a
vast number of proposals in the literature for alternatives, many of them concerned with the classical unification of gravitation and electromagnetism and conceived by
Einstein himself \citep{Goenner2004}. Some of these theories could be ruled out nowadays thanks to the
higher accuracy of measurements in Solar system dynamics and the orbits of spacecrafts. In particular, the BepiColombo mission could provide highly accurate measurements of the parameters $\beta$ and $\gamma$ in the parametrized post-newtonian dynamics formalism. However, other popular theories, such as the Einstein-Cartan model incorporating torsion are still viable \citep{Hehl1976}. Even the pursuit for alternative
theories of gravity, not based upon the principle of equivalence and invariance under
general coordinate transformations, is still investigated. The entropic scenario, based upon the concept of holography, is a very recent enterprise for an alternative
approach to a gravity theory \citep{Verlinde}.

However, it seems unlikely that a novel conception of gravity could emerge merely by abstract thought without the interplay with
crucial experimental results and observations. By following the route of pure abstraction we are necessarily trapped by the traditional ontological
views of previous theories. Some recent astronomical observations have revealed anomalies which, if confirmed, could require such a rethinking on
our current understanding of gravity. In 2004 Krasinsky and Brumberg indicated that the analysis of all available
radiometric measurements of distances between the Earth and the planets, and also the observations of martian landers and orbiters, showed that the
Astronomical Unit is increasing at a rate $15 \pm 4$ meters per century \citep{Krasinsky}. Later on, a more careful analysis by Standish has shown that the secular
rate is closer to $7 \pm 2$ meters per century \citep{Standish}. Anyway, this is by far too large to be explained by the loss of solar mass due to solar wind and
electromagnetic radiation. An explanation based upon tidal friction caused by the bulge produced by Earth gravity on the Sun has been proposed \citep{Miura}.
However, this model has not been validated and the detailed mechanism for this tidal friction is hypothetical. A secular effect on the eccentricity of planetary motions
have been also unveiled by the recent detailed analysis of the Lunar orbit. The 
secular increase of the eccentricity is very small but, however, is clearly within
the range of precision of Lunar laser ranging. This kind of unexplained observations, after discarding any possible conventional explanation, could give rise to an arena where the status of General Relativity as a complete theory of gravity (at least, at the macroscopic level) could be tested. 

In this paper we have assumed that a conventional explanation is not possible and that an extra force term is necessary in order to incorporate this behaviour in the
post-newtonian formalism. Our objective has been to study and discard several force
terms proposed by dimensional analysis. An anomalous extra force field proportional to the rotational angular momentum of the central body and the radial velocity of the
orbiting planet is the most promising one in the parsimonious explanation of the observational data. This could serve as a methodological guideline in the search for
an extension of General Relativity predicting new effects if this observations continue to be supported by more refined tests.

%
% and use \bibitem to create references. Consult the Instructions
% for authors for reference list style.
%

\end{document}